# The orientation of the Amazonian geoglyphs as a clue for their interpretation

## Giulio Magli[a]*

[a] *Department of Mathematics, Politecnico di Milano, Italy. Giulio.Magli@Polimi.it*

Amazonian earthworks – also called geoglyphs – are thousands of man-made earthen structures, mostly of geometrical shape, which progressively emerged from the tropical forest due  to progressive deforestation. They were probably built between the fifth century BC and the end of the first millennium AD, but archaeological investigation on the culture of their builders  is yet at the beginning. Nevertheless,  a ceremonial – rather than practical – function seems likely, at least for those having a very regular shape. In the present paper, simple remote-sensing technique, combined with the methodological approach of modern Archaeoastronomy,  are applied to study for the first time their orientation. The analysis takes in consideration virtually all known squared and rectangular structures for a total of 326 earthworks. The results show without doubts a non-random choice for their orientation and a clear interest of their builders for the annual cycle of the Sun.

## 1 Introduction

In the 1970s, the government of Brazil launched a series of projects promoting the colonization of Amazonia, including the construction of highways across the tropical forest. This triggered the process of  deforestation for farming. As a byproduct, huge "ditch and bank" earthworks  started to emerge when the tropical vegetation was destroyed, especially in the state of Acre (Pärssinen, Schaan, & Ranzi 2009). Many among the Acre earthworks have a geometrical shape: we can see in particular squares, rectangles, circles, but also diamonds and a few octagons. In many cases a network of double-bank straight "roads" appear to connect them (Saunaluoma, Pärssinen & Schaan 2018, Kalliola, et al. 2024,2026). Up today, more than 1200 earthworks have been recognized, and among these, more than one half  have a recognizable – more or less regular – geometrical shape. It is simply impossible to tell how many structures are actually there; as a general indication, geometrical earthworks have been found in the interfluves of the Acre, Iquiri, and Abunã Rivers while in Bolivia mainly oval/circular ditches are known (Shawn 2012). The dimensions of the structures are impressive: ditches can reach 11 meters wide at the top, with depths varying between 1 and 4 meters; the external embankment, which is now between 0.5 and 2 meters high was probably higher before erosion. Some of the monuments are double-ditched, apparently only for aesthetic reasons. The sides are usually not interrupted by entrances, but in several cases an approaching, double-bank straight road (or more than one, up to one for each side of squared structures) can be recognized. Many earthworks are isolated, but there are also concentrations of two or more, for instance a circle and a square linked by a road or two circles of different sizes with a common ditch in a tangential point.

Since their official discovery, occurred with an aerial recognition carried out by Alceu Ranzi in 1988, the Amazonian earthworks have been called "geoglyphs" in analogy with those of the Nasca Valley in Peru', and this terminology has become common. The analogy is, however, not very suitable, because the Nasca figures and lines are etchings obtained by removing pebbles from the desert surface. Moreover, there is no such thing as the collection of zoomorphic geoglyphs of Nasca in Brazil and, in turn, purely geometric figures are rare or absent in Nasca. Therefore, at least in the present authors' view, the Acre earthworks are much more close to the Hopewell earthworks and roads, constructed – mostly in Ohio – by the Hopewell culture (Lepper 2010) .

After discovery, the presence of pottery made immediately clear that the earthworks were pre-Columbian artefacts, As far as the culture responsible for these buildings is concerned, it took a long way for Amazonian archaeology to recognize that the presence of these extended networks of artifacts implies some sort of complex society, of which we are only starting to understand some detail (Pärssinen et al. 2020. In Shahn (2012) words " "A region that was once thought to be unsuitable for the settlements of pre-Columbian peoples was now the setting of ditches, berms, mounds, and roads, features that are more compatible with the occupation of complex societies." A few archaeological excavations have however been carried out, and available data point to the construction of the earthworks between the fifth-third century BC and 1100 AD, although human occupation of some of the sites can be older. (Schaan et al. 2012, Pärssinen 2021), As far as their function is concerned, the debate is ongoing. One could think to funerary enclosures, but - as far as the present author is aware – no burial has been found inside, and actually the material recovered inside the structures is so scarce that it appears that people kept the central parts clean (Saunaluoma, Pärssinen & Schaan 2018) so that one may even think to taboo locations. Some authors however propose a defensive function, but this conflicts with the fact that the shapes are in many cases perfectly geometrical and the ditches are usually inside the walls (similarly to what happens for many of the pre-historic henges in Europe). Further, no sign of defensive tools (such as wooden palisades) has been found. The location of some sites in the highest possible spots with respect to river valleys can actually lead to think at least to guard posts, but also ceremonial and sacred sites usually share higher locations (in either case, this in turn involves a difficult and yet unsolved problem, namely how much forest did exist at the time of construction ).

All in all, a ceremonial function appears likely at least for strictly geometrical structures. A possible way to explore the mind of builders of monuments so deprived of written and contextual information is to analyse the possible presence of astronomical alignments (Magli 2025). In the case of the Hopewell earthworks, in particular, Archaeoastronomy has proven to be a valid tool in understanding more of this interesting and elusive culture (Romain 2000, Hively and Horn 2023, Magli and Lepper 2025). The aim of the present paper is to present - for the first time, at least as far as the author is aware - a scientific investigation on the possible presence of astronomy in the projects of the Acre geoglyphs, using simple but effective remote sensing tecuques. The results here must be considered as preliminary for a series of reasons: earthworks are continuously discovered, their dating is too wide to allow an inspection of possible interest for the stars (due to precession), the possible presence of a close "artificial" horizon created by trees cannot be fully excluded. In spite of all this, as we shall see, what appears to be a clear interest for the annual cycle of the Sun and its agricultural implications emerges, as well as a more symbolic, specific interest for cardinal orientation.

## 2 Materials and methods

In the present paper satellite imagery provided by Google Earth Pro (GEP) will be used. Actually, GEP has already shown its potentialities for remote sensing archaeological investigations in general (Parcak et al 2016) and Archaeoastronomy in particular (Magli 2018). Specifically, GEP is well suited for archaeoastronomy due to the impressive accuracy of geographical north, and therefore in measuring azimuths, while importing satellite images in processing programs (e.g., in AutoCAD or ArcGIS) may introduce additional errors (Potere 2008).Another useful tool is the historical archive of the program which, in some cases, allows to measure bearings of monuments which are covered by clouds or depicted in unfavourable conditions in more recent images. GEP coverage of the Acre state is generally speaking very good, with resolutions usually of 30 cm. Thus, errors in azimuths arise mostly as human errors in individuating the sides of the earthworks. Overall, we can prudentially estimate the azimuths data presented to have a uncertainty of ±1°, a value which is by far sufficient for our purposes. The collected data are given in the supplementary material; the progressive number correspond to the numbering of the placemarks in the kmz file (also available with the paper).

The data have been processed using standard statistical tools: a mixture of Gaussian kernels distributions, which helps to individuate peaks. The significance of these peaks has been assessed applying standard p-value test of statistical significance with respect to the null hypothesis. Low p-values (usually of the order of 0.05 or less) indicate greater statistical significance.

To collect the data I have used existing databases kindly provided by J.Q. Jacobs (http://www.jqjacobs.net/archaeology/geoglyph.html) and Kalliola and Pärssinen (https://doi.org/10.23729/da98421d-d65d-4045-bc93-344a0837cc93). I located each structure on GEP, searching for each one the best available image in the historical archive of the program. Then, the following procedure was applied:

- All square and rectangular structures have been considered (see Fig 1,a) , including a few which have rounded corners
- All parallelogram structures with internal angles deviating not more than 2 degrees from 90 degrees have been included (see Fig 1,b)
- All other structures have been excluded (see Fig 1,c)

At the end of the procedure, of about 1300 known earthworks, 324 have been selected as satisfying the above criteria. Then, the following definition of azimuth of a rectangular earthwork has been applied: the azimuth is, between the azimuths of two adjacent sides, the one closer  to due east (in case of slightly bended parallelograms, a average has been taken). This procedure is analogue to that used for Roman “orthogonal” towns (Magli 2008, Rodríguez-Antón, González-García, & Belmonte 2018), and is specifically suited for archaeoastronomical investigations. The result is in fact a distribution of data between 45° and 135° (for instance if a side has a bearing of 40°,  the adjacent  side will return 130° and this is the azimuth which will be considered). Within the arc 45°/135° a smaller arc is defined by the rising extrema of the Sun (with a flat horizon). This arc is fundamental for the present paper; it of course depends pointwise from the latitude, but for the area of study we can assume an average latitude of 9° 30’ S and a solar arc of 66°/114°. Another possibly important solar phenomenon is that of the zenith passages. Again assuming an average latitude of 9° 30’, these passages occur with azimuth at rising around 99° 30’ (dates 23 February/18 October). Finally, the Moon northernmost extrema correspond to azimuths 61°/119°.

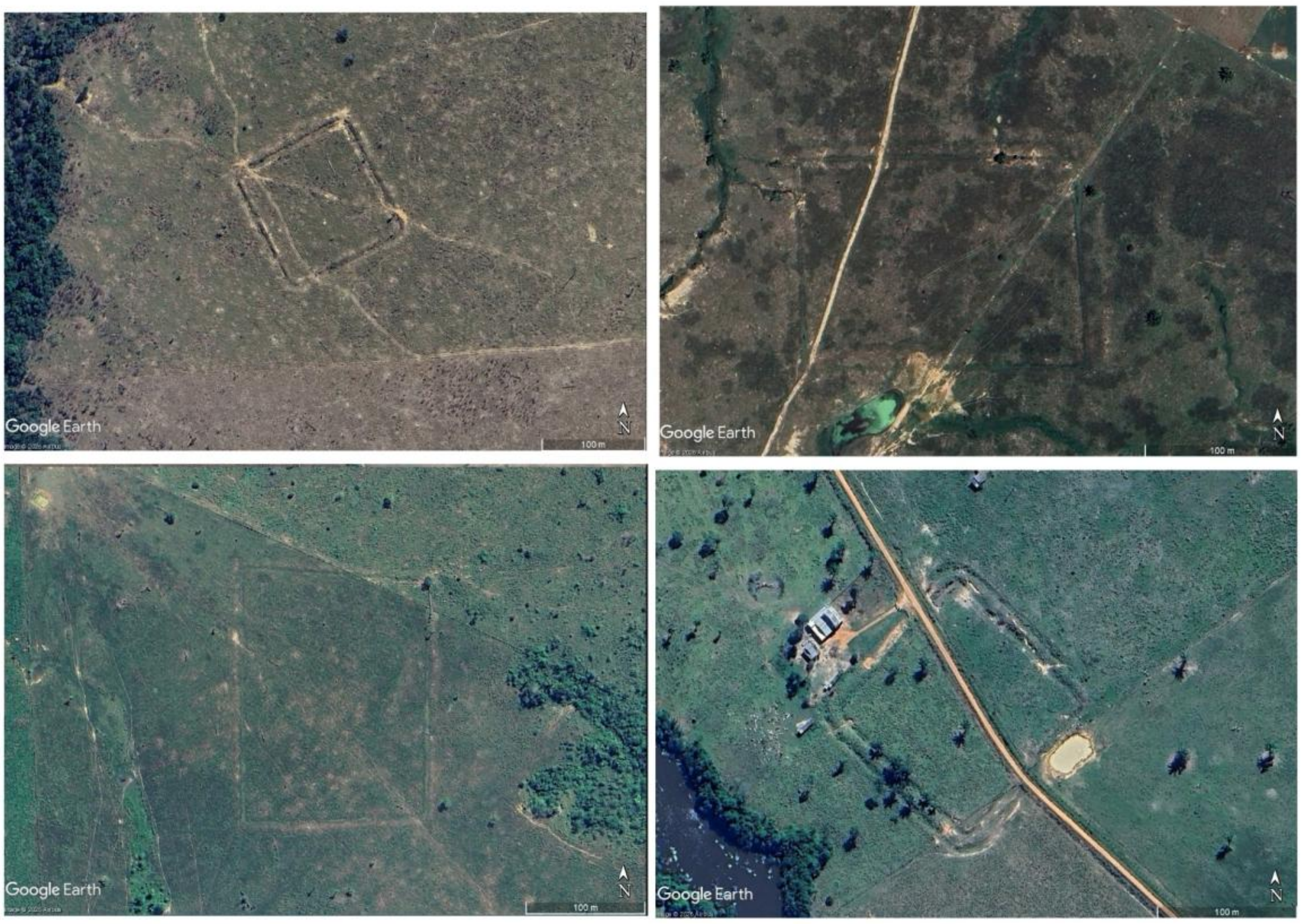

**Figure 1**

a) example of square b) example of parallelogram included in the database c) example of parallelogram excluded from the database d) example of cardinally oriented rhombus (images courtesy Google Earth Pro, editing by the author)

## 3 Results

The azimuths of the structures are reported in the histogram of Fig. 2. It is evident already at sight that the distribution is far from being uniform.

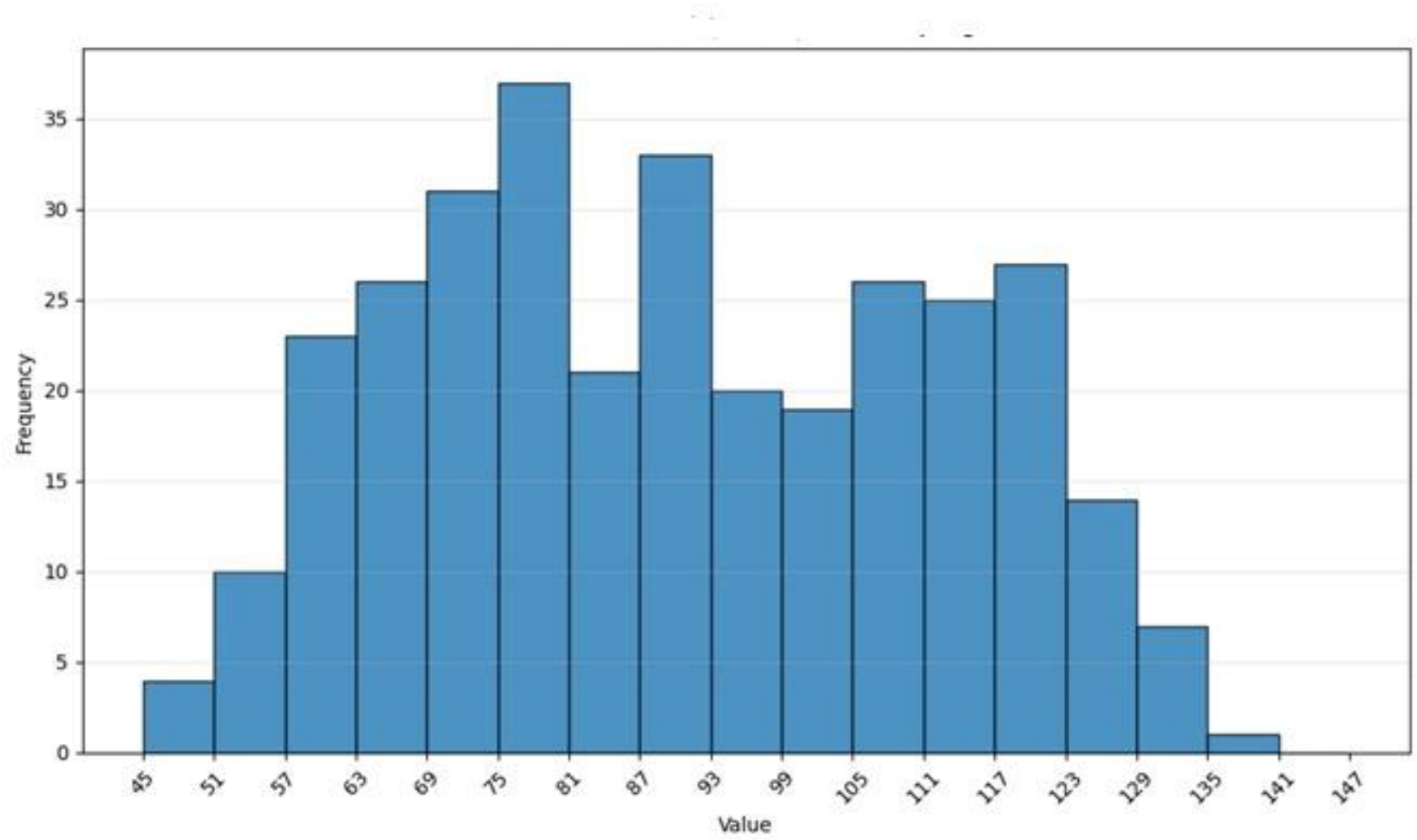

**Figure 2** Data Histogram

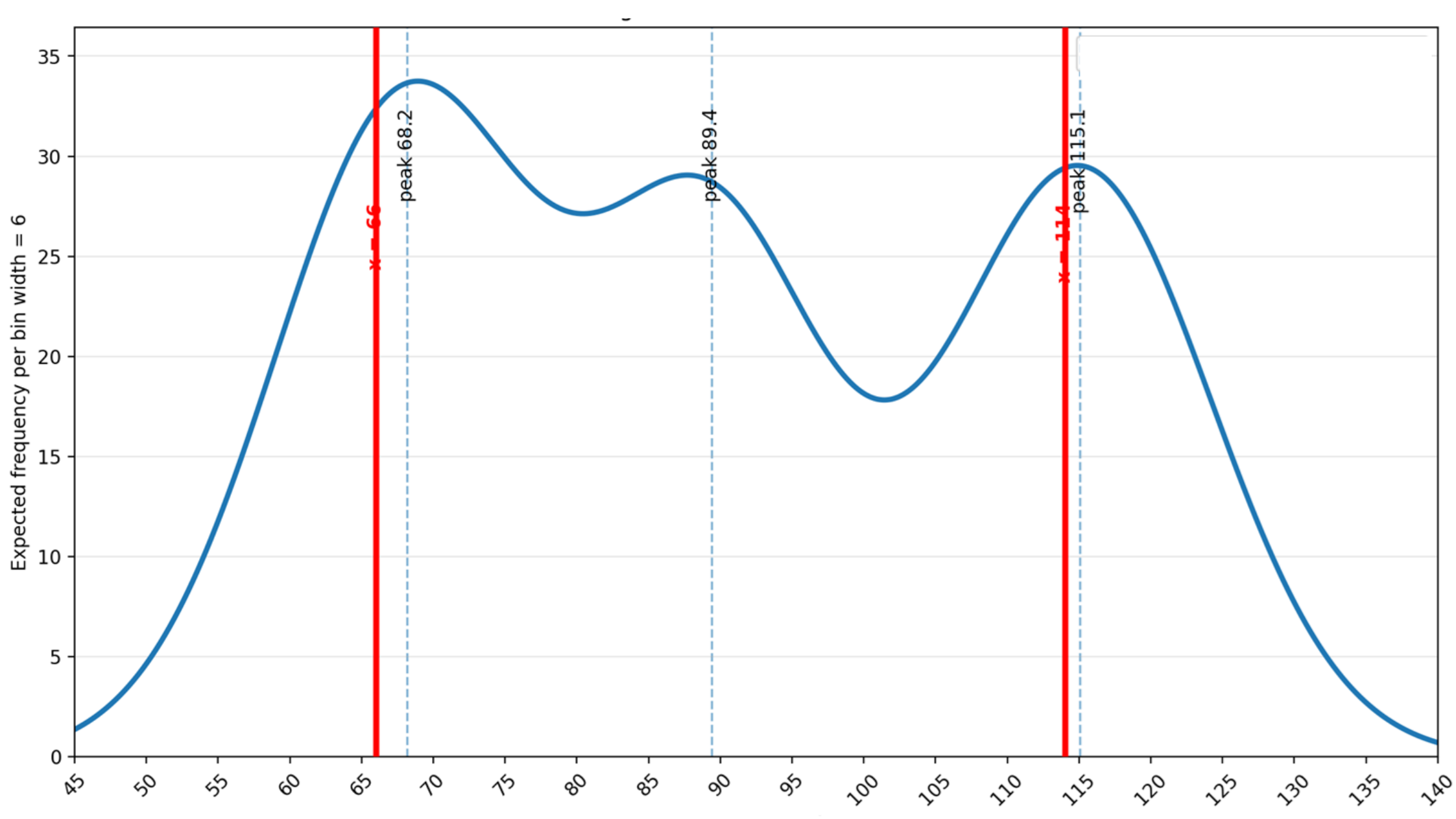

**Figure 3** Curvigram of the data expressed as a 3-Gaussian mixture plot. Red vertical lines indicate the winter and the summer solstice (average) rising azimuths with a flat horizon

This is confirmed by a p-value test on the whole data which returns $1.5 \times 10^{-15}$. Therefore, our first result is that the orientation of the Acre earthworks is not random but due to explicit design. Of course, this design may be due to practical reasons, but going in more depth we can see that it was very likely astronomical. In fact, applying the Gaussian kernels method mentioned above (but the result are robust with respect to the choice of the model for the kernels, for instance the results with Epanechnikov kernels do not change significantly) we get the "curvigram" given in Fig. 3 (usually, these graphs are plotted with the relative frequency on the y-axis; for clarity, I prefer here to use simply the number of instances). The azimuths of the solstices are drawn in red for reference.

It is apparent the general fact that the builders were interested in the solar arc, as shown by the rapid decay of the data outside the arc. The distribution has three statistically significant peaks at ≈ 68°, 89,5° , and 115°. To be precise, there is strong statistical evidence for the two peaks at 68° and 115° ($p \approx 5 \times 10^{-10}$) and borderline evidence ( $p \approx 0.5$ ) for the central one. As a matter of fact, however, three out of three have an astronomical significance. Indeed,

- the *less* pronounced peak is that at 89,5°. The equinoxes can of course be mentioned, but the notion of *cardinal orientation* has to be preferred, since it is much more likely that ancient population were interested in the quadripartition of the world rather than in the midpoint of the apparent motion of the Sun (although of course the two things do not exclude each other). On the significance of this peak see discussion below

- the peak at 115° is a clear indication of interest in the summer solstice sunrise

- the peak at 68° is not far from winter solstice sunrise, but a difference of two degrees in azimuth implies a relevant difference in dates, pointing to dates around 29 may/14 July (increased azimuths due to horizon heights – trees in this

case - might occur considering setting but this seems unlikely). A proposal of explanation will be given below

**4 Discussion**

The non-randomness of the data, together with the interest in the solar arc are a first, although rough, indication that astronomy was used in the project of the Arce earthworks. A calendrical (seasonal) interest in the annual cycle of the Sun is the likely explanation for the summer solstice peak. Less clear is the peak at 68° which corresponds to two dates not far from the winter solstice. However, one possibility to interpret it is the following. Acre has a long rainy season which starts in October and drops to an end in a relatively rapid way at the end of May. For instance in the capital Rio Bravo an average of about 100 mm of rain in May drops to 45 mm in June. Therefore the Sun of the end of May was the harbinger of the dry season, a role which would have been unsuitable for the winter solstice Sun some 22 days later. Furthermore, at least today, a double harvesting of Corn is made. One lasts from September to January/February, and the other begins immediately afterwards, with harvesting at the end of May.

Finally, as far as the cardinal peak is concerned, it is worth recalling that orientation of sacred structures to the cardinal points has been a constant among many civilizations (see e.g. Magli 2025 and references therein). In the present case, its existence can perhaps be reinforced by the following consideration. Among the structures which have been excluded from the present analysis there is a certain number with a almost perfect rhomboidal shape, in many cases doubled (see Fig.1,d). It is difficult to imagine which kind of practical purpose an object of this kind may have had, so that they are likely ceremonial. At present, their number is not enough to allow a statistical assessment, but there is the distinct possibility that their *diagonals* were intentionally oriented, and in particular, for at least eight of them the diagonals are oriented to the cardinal points (see again Fig. 1,d).

All in all, the results of the present paper point towards the interpretation of the Acre geoglyphs – at least in the case of rectangular shapes – as ceremonial structures, built by a culture interested in following the annual cycle of the Sun. The work has an intrinsic limitation, though, due to the fact that many earthworks – we do not know how many – are covered by the surviving forest. Non invasive airborne Lidar techniques combined with focussed archaeological excavations will be the best way to enlarge our knowledge on these fascinating, unexpected structures in the future.

**References**

Kalliola, R., Pärssinen, M., Ranzi, A., Seppa, I., & Barbosa, A. D 2024 Geography of ancient geometric earthworks and their builders in southwestern Amazonia Acta Amaz. 54 (4) • Oct-Dec 2024

Kalliola, R., Pärssinen, M., Ranzi, A., & Barbosa, A. D. (2026). Ancient Amazonian Earthwork Roads: Unveiling Ceremonial, Livelihood, and Networking Significances. *Latin American Antiquity*, 1-16.

Hively, R., & Horn, R. (2023). More Topography, Astronomy, and Geometry at the Newark Earthworks. *Midcontinental Journal of Archaeology*, *48*(1), 1-45.

Lepper, B. T. (2010). The ceremonial landscape of the Newark Earthworks and the Raccoon Creek Valley. *Hopewell settlement patterns, subsistence, and symbolic landscapes*, 97-127.

Magli, G. (2008). On the orientation of Roman towns in Italy. *Oxford Journal of Archaeology*, *27*(1), 63-71.

Magli, G. Astronomy and Feng Shui in the projects of the Tang, Ming and Qing royal mausoleums: A satellite imagery approach. Arch. Res. Asia 2018, 17, 98–108.

Magli, G. (2025). *Archaeoastronomy: introduction to the science of stars and stones*. Springer Nature.

Magli, G., & Lepper, B. (2025). Going straight in a sacred landscape: The Great Hopewell Road. *Studies in Digital Heritage*, *9*(1), 37-54.

Parcak, S.; Gathings, D.; Childs, C.; Cline, G.M.E. Satellite Evidence of Archaeological Site Looting in Egypt: 2002–2013. Antiquity 2016, 90, 188–205.

Pärssinen, M., Schaan, D., & Ranzi, A. (2009). Pre-Columbian geometric earthworks in the upper Purús: a complex society in western Amazonia. *Antiquity*, *83*(322), 1084-1095.

Pärssinen, M., Balée, W., Ranzi, A., & Barbosa, A. (2020). The geoglyph sites of Acre, Brazil: 10 000-year-old land-use practices and climate change in Amazonia. *Antiquity*, *94*(378), 1538-1556.

Pärssinen, M. (2021). Tequinho Geoglyph Site and Early Polychrome Horizon 300 BC-AD 300/500 in the Brazilian State of Acre. *Amazonica: revista de antropologia*, *13*(1), 177-220.

Potere, D. Horizontal Positional Accuracy of Google Earth's High-Resolution Imagery Archive. Sensors 2008, 8, 7973–7981

Rodríguez-Antón, A., González-García, A. C., & Belmonte, J. A. (2018). Astronomy in Roman urbanism: a statistical analysis of the orientation of Roman towns in the Iberian Peninsula. Journal for the History of Astronomy, 49(3), 363-387.

Romain, W. F. (2000). *Mysteries of the Hopewell: astronomers, geometers, and magicians of the Eastern Woodlands*. University of Akron Press.

Schaan, D. 2012. *Sacred Geographies of Ancient Amazonia: Historical Ecology of Social Complexity* Left Coast Press, Walnut Creek, 233p.

Schaan, D., Pärssinen, M., Saunaluoma, S., Ranzi, A., Bueno, M., & Barbosa, A. (2012). New radiometric dates for precolumbian (2000–700 BP) earthworks in western Amazonia, Brazil. *Journal of Field Archaeology*, *37*(2), 132-142.

Saunaluoma, S., Pärssinen, M., & Schaan, D. (2018). Diversity of pre-colonial earthworks in the Brazilian state of Acre, southwestern Amazonia. *Journal of Field Archaeology*, *43*(5), 362-379.